\documentclass[aip,amsmath,amssymb,reprint,jap,superscriptaddress]{revtex4-1}

\usepackage{graphicx}
\usepackage{dcolumn}
\usepackage{bm}

\usepackage[utf8]{inputenc}
\usepackage[T1]{fontenc}
\usepackage{mathptmx}
\usepackage{etoolbox}

\usepackage{xcolor}
\usepackage{epsfig}
\usepackage{bm}

\usepackage[utf8]{inputenc}
\usepackage[T1]{fontenc}
\usepackage{mathptmx}
\usepackage{etoolbox}
\usepackage{soul}
\makeatletter
\def\@email#1#2{%
 \endgroup
 \patchcmd{\titleblock@produce}
  {\frontmatter@RRAPformat}
  {\frontmatter@RRAPformat{\produce@RRAP{*#1\href{mailto:#2}{#2}}}\frontmatter@RRAPformat}
  {}{}
}%
\makeatother
\begin{document}

\preprint{AIP/123-QED}

\title{A Molecular Dynamics Study of Laser-Excited Gold}
\author{Jacob M. Molina}
 \email{jmmolina@nevada.unr.edu}
 \affiliation{University of Nevada, Reno}
\author{T. G. White}%
 \affiliation{University of Nevada, Reno}
\date{\today}
\begin{abstract}
The structural evolution of laser-excited systems of gold has previously been measured through ultrafast MeV electron diffraction. However, there has been a long-standing inability of atomistic simulations to provide a consistent picture of the melt process, concluding in large discrepancies between the predicted threshold energy density for complete melt, as well as the transition between heterogeneous and homogeneous melting. We make use of two-temperature classical molecular dynamics simulations utilizing three highly successful interatomic potentials and reproduce electron diffraction data presented by Mo \emph{et al}. We recreate the experimental electron diffraction data employing both a constant and temperature-dependent electron-ion equilibration rate. In all cases we are able to match time-resolved electron diffraction data, and find consistency between atomistic simulations and experiments, only by allowing laser energy to be transported away from the interaction region. This additional energy-loss pathway, which scales strongly with laser fluence, we attribute to hot electrons leaving the target on a timescale commensurate with melting.  
\end{abstract}

\maketitle

\section{Introduction}

Ultrafast laser excitation of metals is able to bring material into a state far from equilibrium \cite{Eq1,Eq2}. The preferential and rapid heating of one subsystem over the other leads to a system of highly coupled cold ions immersed in a partially degenerate electron sea \cite{Ichimaru}. These transient states commonly occur during the formation of high energy density plasmas, including warm dense matter (WDM), with particular relevance to laser micromachining \cite{WDM1, WDM2,WDM3, WDM4} and inertial confinement fusion experiments \cite{Fusion}. In the laboratory, these transient states serve as a testbed where quantum mechanical theories of electron-ion interactions, nuclei dynamics, and phase transitions can be validated \cite{MoPhase,Mo2018,Szymon2013,Ernstorfer2009}. 

Historically, the response of the electron subsystem has been measured in optical pump-probe experiments \cite{Probe1,Probe2,Probe3,Probe4,HolstValid}. However, these model-dependent techniques provided only a surface measurement of the electron properties and limited information on the ionic response. More recently, X-ray scattering experiments have measured the bulk electron temperature by observing the electron plasmon feature \cite{Fletcher}. In contrast, the bulk ion temperature has only been inferred from the atomic structure, measured through ultrafast electron \cite{Mo2018} or X-ray diffraction \cite{DW1, DW2}. When the ionic system is still crystalline, the Debye-Waller factor (i.e., the reduction in the intensity of the Laue diffraction peaks) is a practical, albeit model-dependent, method \cite{DW1, DW2}. However, modeling the decay of the Laue diffraction peak, which is ultimately dependent on the root-mean-square (RMS) deviation of the atoms from their lattice positions, is dependent on several physical parameters:
\begin{itemize}
    \item The energy density of the sample - $\epsilon$
    \item The electron-ion equilibration rate - $g_{ei}(T_e, T_i)$
    \item The electronic heat capacity - $C_{e}(T_{e})$
    \item The ionic heat capacity - $C_{i}(T_{i})$
    \item The Debye temperature - $T_{D}(T_e, T_i)$
\end{itemize} 
where the generally assumed dependence on electron temperature ($T_e$) and ion temperature ($T_i$) is given. While the electronic and ionic heat capacities are well constrained for gold \cite{Holst2018,Lin2008}, the remaining three parameters exhibit large uncertainties in the literature. In particular, theoretical predictions of the equilibration rate vary by up to an order of magnitude \cite{Lin2008,Med2020,Brown2016,Smirnov2020,Migdal2013,Wang1994,Papaconstantopoulos2015,Petrov2013}. 

During the last decade, three different experiments have used ultrafast electron diffraction to measure the decay of the Laue diffraction peaks in laser-irradiated gold. They each attempt to elucidate the effects of non-equilibrium species on the interatomic potential - i.e., measure the non-equilibrium Debye temperature and answer the long-standing question surrounding the existence of bond hardening/softening in warm dense gold \cite{Mo2018,Szymon2013,Ernstorfer2009,Recoules}. For comparable fluences, each experiment measures similar decays in the Laue diffraction peaks. However, they reach opposing conclusions; this is primarily due to different assumptions regarding the behavior of the electron-ion equilibration rate and the initial energy density deposited into the sample. For example, in the work of Daraszewicz \emph{et al} the assumed initial energy density, $\mathcal{E}$, calculated by taking into consideration purely the reflected and transmitted light, was reduced by a factor of $\eta$ \cite{Szymon2013},
\begin{equation}
         \epsilon= \eta \mathcal{E},
\end{equation}
where $\epsilon$ is the corrected energy density. They found a value of $\eta\sim0.5$, measured through independent optical absorption methods at lower laser fluences \cite{SzymonEta}. At higher fluences, they found it necessary to treat $\eta$ as a free parameter. This extra energy-loss pathway, specific to the target and experimental geometry, was attributed to ballistic/fast electron transport away from the system - this, despite contemporary claims of the negligibly of $\eta$ \cite{ChenEta}. 

One method of modeling such experiments is classical molecular dynamics (MD). When coupled to an electronic system through an appropriate thermostat, such simulations can, in one dimension, simulate the full target thickness; thus, capturing the necessary strongly coupled nature of the ions \cite{LangDyamics}. However, recent efforts using the highly optimized and validated interatomic potential developed by Sheng \textit{et al} \cite{Sheng2011, Chen2018} were unable to reproduce the experimental results for all energy densities in tandem \cite{ZengMD}. 

In contrast to previous simulation work, we treat both the electron-ion equilibration rate ($g_{ei}$) and initial energy density ($\epsilon$) as free parameters to form a consistent description of melting in warm dense gold. Initially, we make use of a constant $g_{ei}$ throughout each simulation, as well as exploring three different interatomic potentials. In all cases, we are able to obtain excellent agreement between the decay of Laue diffraction peaks obtained experimentally and those calculated from synthetic diffraction patterns. We find a strongly energy-dependent electron-ion equilibration rate in addition to considerable differences between the assumed energy density, $\mathcal{E}$, and the energy density in the simulations, $\epsilon$. In fact, for each interatomic potential that we have tested, we found that the success of the model is contingent on allowing energy to escape the target region. This additional energy loss, which we also characterized by $\eta$, scales strongly with laser intensity. Additionally, we have implemented several published electron temperature-dependent and ion temperature-dependent $g_{ei}$ models \cite{Holst2018,Lin2008,Med2020,Smirnov2020,Migdal2013,Petrov2013}. For all forms of $g_{ei}$, consistency between the simulated and measured diffraction patterns is only achieved by reducing the absorbed energy density. 

By introducing the free parameter $\eta$, we can form a consistent description of the melting process in warm dense gold, resolving long-standing discrepancies between MD simulations and experiments. Our finding of a non-negligible $\eta$ parameter is in good agreement with the work of Daraszewicz \textit{et al}; however, we additionally observe a strong dependence on laser intensity.

\section{Computational Methods}
 We perform large-scale MD simulations in the LAMMPS software package \cite{lammpscitation} in the canonical ensemble. The ions, treated explicitly, are coupled to an electron sub-system through a Langevin thermostat within the two-temperature model (TTM) \cite{Rutherford2007,Duffy2006}. Justified by the rapid electron thermalization time\cite{Fann1992}, we heat the electrons instantaneously to simulate the laser-matter interaction; we verify the accuracy of this procedure in the supplementary information by comparing the results of the forthcoming analysis with simulations taking the temporal width of the laser pulse into account. Synthetic diffraction patterns are produced in order to compare with experimental electron diffraction data of Mo \emph{et al} \cite{Mo2018}.

 In this work, we focus on benchmarking the data obtained by Mo \textit{et al}, which is expected to be below the threshold for considerable changes in interatomic bond-strength \cite{Szymon2013}. In keeping with the experimental conditions, we model free-standing 35-nm single-crystalline gold films and compare our Laue peak decay to measured results for the three assumed energy densities $\mathcal{E}$=0.18~MJ/kg, $\mathcal{E}$=0.36~MJ/kg, and $\mathcal{E}$=1.17 MJ/kg, referred to here as the low, intermediate, and high energy density cases.

We model the non-equilibrium conditions of the laser-irradiated system via a traditional TTM that describes the evolution of the electron temperature through,

\begin{equation}
    C_{e}(T_{e}) \frac{\partial T_{e}}{\partial t}=-g_{ei}(T_{e}-T_{i}),
\end{equation}
where $C_{e}(T_{e})$ is the temperature-dependent electron heat capacity. All simulations utilize an electronic heat capacity derived from \emph{ab-initio} simulations \cite{Holst2018}, and found to be in good agreement with prior thermal conductivity experiments \cite{HolstValid}. Within each simulation, the electron-ion equilibration rate is treated as a constant and we employ a 1~fs timestep throughout. In addition, prior to coupling the two sub-systems, we equilibrate just the ions for 6.5~ps using a velocity-scaling thermostat and Berendsen barostat \cite{Beren} in order to bring the system to a temperature of 300~K at atmospheric pressure. This introduces an initial energy density into the ionic sub-system of $\sim 0.05$~MJ/kg. 

The ballistic electrons, accelerated by the optical laser, have a mean free path of $\sim$100~nm, approximately three times the thickness of the foil; thus, we approximate the heating as isochoric \cite{BallisticRange1,BallisticRange2,BallElecTrans,Rev2Ballistic}. The high thermal conductivity \cite{Conduct} also allows us to retain a single spatially invariant electron temperature, and thus avoid complications due to lattice expansion \cite{ZengMD}. This is particularly important to avoid spurious edge effects which may initiate heterogeneous melting.  For the duration of the simulation, to model the electron-ion energy exchange, we couple the ions to this single-temperature electron bath via a Langevin thermostat given by \cite{LangevinPaper,Mabey},
\begin{equation}
    m\frac{\partial \textbf{v}_{i}}{\partial t}=\textbf{F}_{i} - \gamma \textbf{v}_{i} + \textbf{f}_{L}(T_{t}),
\end{equation}
where  $\textbf{v}_{i}$ is the velocity of particle $i$ and $\textbf{F}_{i}$ is the force acting on atom $i$ due to the interaction with the surrounding atoms at time $t$. Here, $m$ is the mass of the ions, $k_{b}$ is Boltzmann's constant, and $\gamma$ is the friction parameter that characterizes the electron-ion equilibration rate. The friction parameter is related to the equilibration rate through $\gamma =g_{ei} m/3nk_{b}$, where $n$ is the number density of atoms \cite{Rutherford2007,Duffy2006}. The $\textbf{f}_{L}(T_{t})$ term is a stochastic force term with a Gaussian distribution and a mean and variance given by \cite{Rutherford2007,Stochastic},
\begin{equation}
    \langle \textbf{f}_{L}(T_{t}) \rangle = 0,
\end{equation}
\begin{equation}
    \langle \textbf{f}_{L}(T_{t}) \cdot \textbf{f}_{L}(T_{t'}) \rangle = 2\gamma k_{b} T_{e} \delta(t-t').
\end{equation}

\begin{table}[h]
\caption{\label{tab:sheng} Experimentally measured properties of Face-Centered Cubic (FCC) gold alongside those reproduced in molecular dynamics simulations using the interatomic potentials developed by Sheng \textit{et al} and the electron temperature-dependent (ETD) potential by Norman \textit{et al}. Here, the values depicted for the Norman \textit{et al} potential were for the variant of the potential calculated for $T_{e}\sim$ 0.1~eV.}
\begin{ruledtabular}
 \begin{tabular}{c c c c}       
Property & Sheng\footnote{Reference \cite{Sheng2011}} & Norman\footnote{Reference \cite{HighTPot}} & Expt.\\
\hline
Melt Temperature~(K) & 1320 & 1208 & 1337\footnote{Reference \cite{MeltTemp}}\\
Lattice Constant~({\AA}) & 4.078 & 4.155 & 4.078\footnote{Reference \cite{LattConst}}\\
Liquid Density~(g/cm$^{3}$)& 17.10 & 17.70& 17.10\footnote{Reference \cite{LiqDen}}\\
Melting Enthalpy~(KJ/mol)& 11.10 & & 12.80\footnote{Reference \cite{EnthalpyExp}}\\
\end{tabular}                   
\end{ruledtabular}
\end{table}

\begin{figure}[t]
  \centering
  \includegraphics[width=\linewidth]{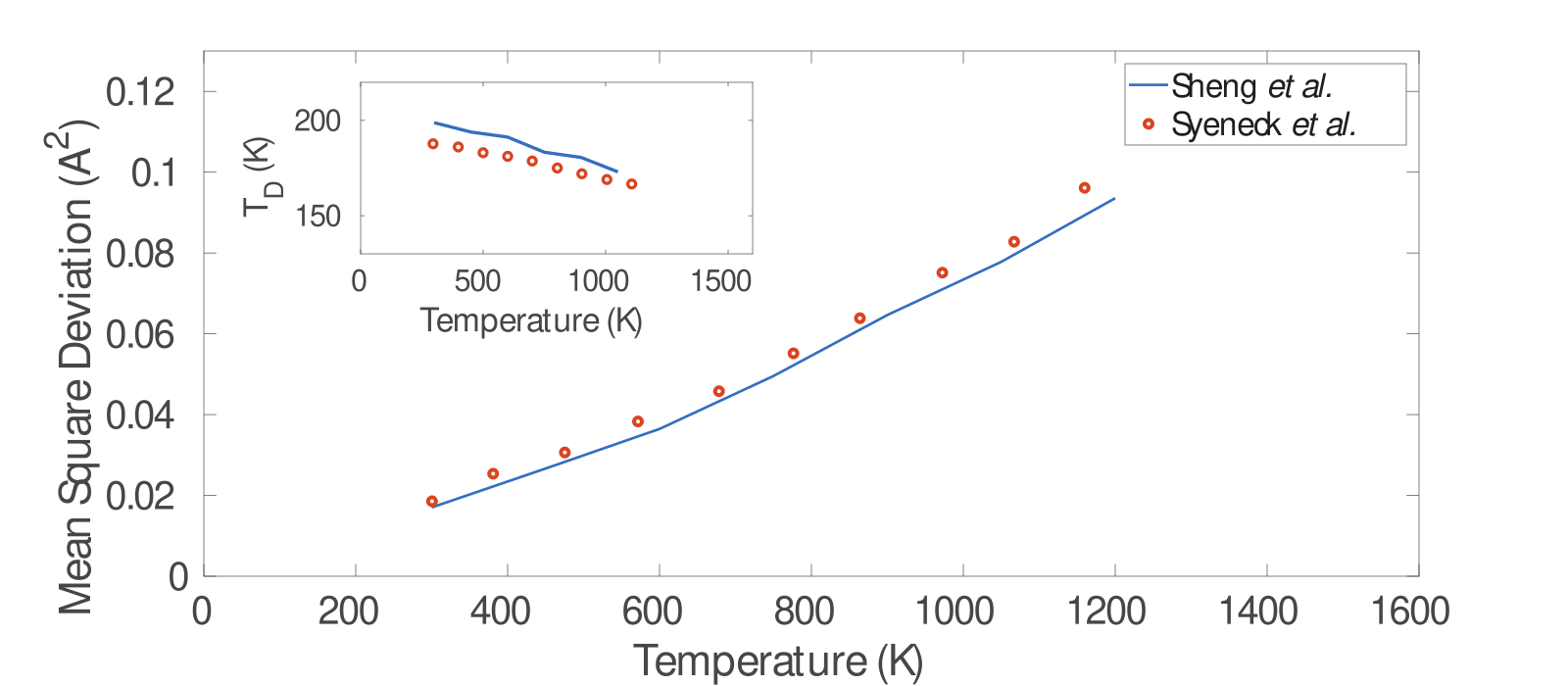}
  \caption{Comparison of the atomic mean square deviation and Debye temperature, $T_{D}$, produced by interatomic potential developed by Sheng \textit{et al} \cite{Sheng2011} and the values obtained from X-ray diffraction measurements at equilibrium temperatures up to melt by Syneček \textit{et al} \cite{DebyeExp}. }
  \label{fig:DebyeT}
\end{figure}

We utilize the highly optimized embedded atom method (EAM) interatomic potential developed by Sheng \emph{et al} \cite{Sheng2011}. Under equilibrium conditions, the properties of the Sheng potential are closely matched with experimental values, as shown in Table \ref{tab:sheng}. Additionally, as demonstrated in Fig. \ref{fig:DebyeT}, the atomic RMS deviation and corresponding Debye temperature, obtained in X-ray diffraction experiments, are well matched at equilibrium temperatures up to melt. In non-equilibrium matter, the Sheng potential has been able to successfully match experimental lattice disassembly times after laser irradiation \cite{Chen2018}.  

We utilize two simulation geometries, either a parallelepiped of (86$a_{0}$ $\times$ 15$a_{0}$ $\times$ 15$a_{0}$), or (86$a_{0}$ $\times$ 70$a_{0}$ $\times$ 70$a_{0}$), where $a_{0}=4.078~${\AA} is the lattice constant of gold predicted by the Sheng potential. The smaller volume contained 77,400 atoms, whereas the larger volume contained 1,685,600 atoms. In the larger direction, which corresponds to the 35~nm thickness of the gold foils used in the experiment, the simulation box is much larger than the size of the target geometry, creating a front and rear surface capable of expansion. In the two smaller directions, we utilize periodic boundary conditions. The smaller geometry is employed when calculating the decay of the Laue peaks. The larger volume, which produces the same Laue decay curve as the smaller volume, allows for a higher resolution in reciprocal space and is used to create the synthetic spatially-dependent diffraction patterns out to $50$~ps.

\section{Matching Experimental Data}
\begin{figure*}[t!]
    \centering
    \includegraphics[width=0.85\linewidth]{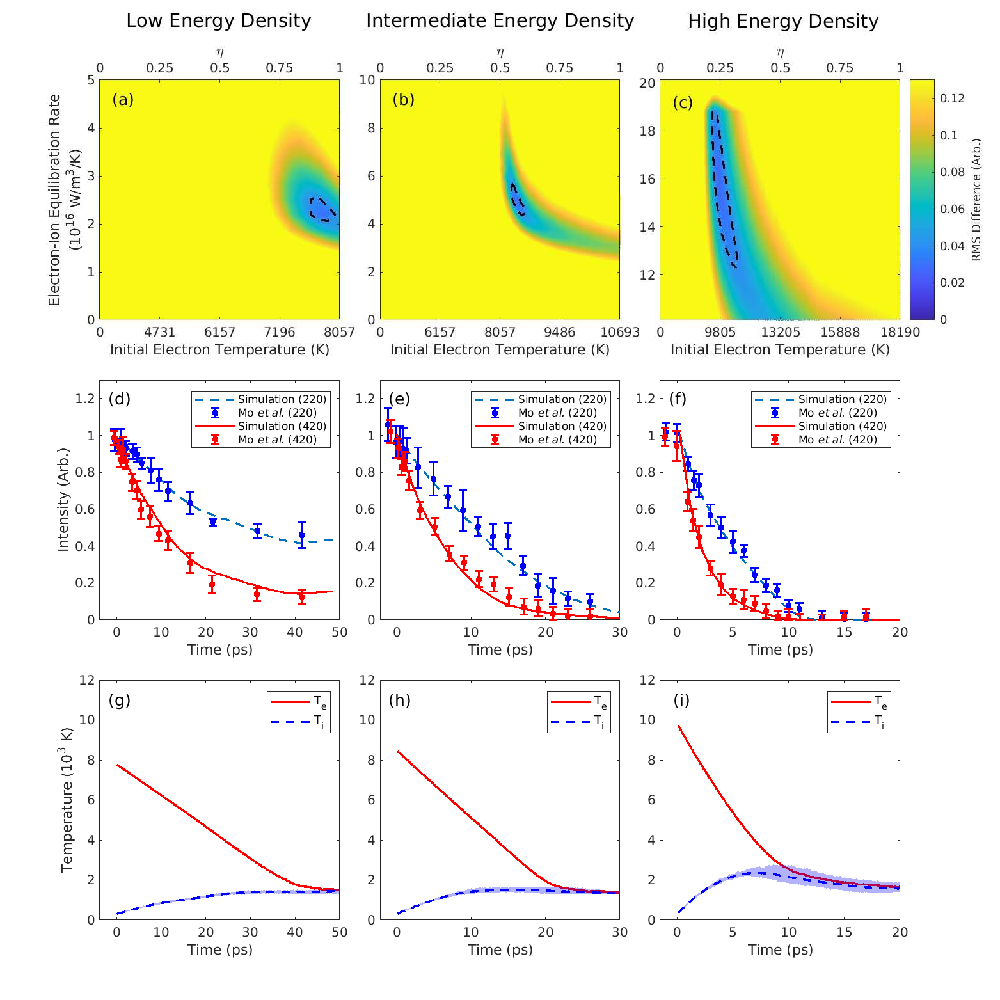}
    \caption{Comparison between experimentally obtained Laue decay curves \cite{Mo2018} and those produced through molecular dynamics simulations. (a-c) The RMS difference between simulated and experimental decay curves for a range of initial electron temperatures and electron-ion equilibration rates. Here, the RMS difference of the (220) and (420) have been added in quadrature. The area enclosed by the dashed line encompasses the region in which the simulation results lie within the experimental error bars.(d-f) A comparison, for each energy density, of the experimental and simulated decay of the (220) and (420) Laue peaks for best fit values of $T_{e}^{0}$ and $g_{ei}$ obtained in panels (a-c). (g-i) The corresponding temporal evolution of the electron and ion temperatures. Here, the shaded area in blue represents in the variation in the ion temperature across the sample.}
    \label{fig:constanteq}
\end{figure*} 
We use the decay of the (220) and (420) Laue diffraction peaks as a metric to compare the simulation and experiment. We obtain the intensity of the diffraction peaks from the static structure factor of the system, which is easily calculated from the Fourier Transform of the atomic positions \cite{ssf1, ssf2}. For 3.2~MeV electrons, whose wave vector ($\sim2\times10^{13}$~rad/m) is orders of magnitude higher than their scattering vector ($\sim10^{10}$~rad/m), the Ewald sphere can be considered flat over a finite region of reciprocal space. As such, we take the $Q_x$-$Q_y$ slice of reciprocal space, where the $x$ and $y$ coordinates run parallel to the surface of the foil. The scattering intensity is calculated by integrating over a specific reflection.

In finding the best fits for all three experimental data sets, we performed hundreds of simulations utilizing the small simulation geometry. We considered initial electron temperatures ($T_{e}^{0}$) which, when converted into energy densities via the electron heat capacity \cite{Holst2018}, correspond to a range in $\eta$ of $0 \leq \eta \leq 1$, with a step size of $\Delta T_{e}^{0}=100$~K. We considered values of $g_{ei}$ in the range of $0 \leq g_{ei} \leq 20\times 10^{16}$~W/m$^{3}$/K with a step size of approximately $\Delta g_{ei} \sim 0.3 \times 10^{16}$~W/m$^{3}$/K. We quantify the best fit by identifying which ($T_{e}^{0}, g_{ei}$) pair minimizes the RMS difference between the experimental and calculated decay in the intensity of the (220) and (420) Laue diffraction peaks. For each of the three energy densities investigated, this difference is shown in Figs. \ref{fig:constanteq}(a-c), with the corresponding best decay curve and temperature evolution for the sub-systems given in Figs. \ref{fig:constanteq}(d-i). The black dashed line in Figs. \ref{fig:constanteq}(a-c) encloses the region in which the simulation results lie within the experimental error bars. In each case, we are able to find a ($T_{e}^{0}, g_{ei}$) pair which gives excellent agreement with the experimentally measured decay curves. 

\begin{figure*}[tb!]
    \centering
    \includegraphics[width=0.7\linewidth]{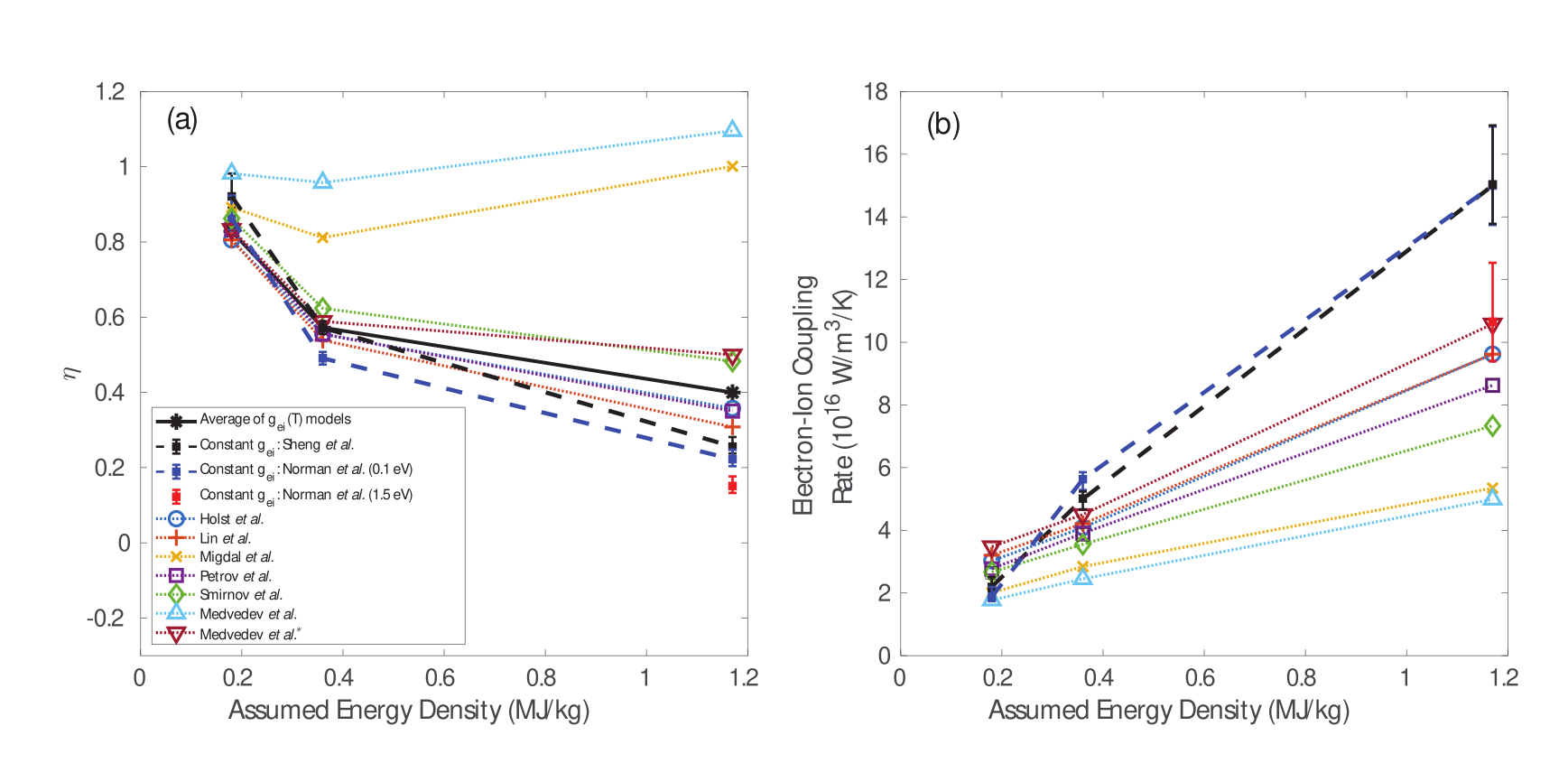}

    \caption{(a) Evolution of best fit values of $\eta$ for several functional forms of the electron-ion equilibration rate\cite{Holst2018,Lin2008,Med2020,Smirnov2020,Migdal2013,Petrov2013}. Here, Medevdev \textit{et al}* is used to denote the $g_{ei}(T_e,T_i)$ prediction that is both electron and ion temperature-dependent. In our calculations of the average $\eta$ for all theoretical models, we exclude the outlier results produced by the solely electron temperature-dependent model calculated by Medevdev \textit{et al} at a constant ion temperature of 300~K, and the Migdal \textit{et al} model. (b) The average value of $g_{ei}$ for a given energy density up until melt. The value plotted was calculated by taking the average over the $g_{ei}$ value present out until 50~ps, 25~ps, and 10~ps in the low, intermediate, and high energy density cases, respectively.}
    \label{fig:finalresult}
\end{figure*}

For the low energy density case, we find that values of $T_{e}^{0}=7800\pm300$~K and $g_{ei}=2.2\pm0.5\times10^{16}$~W/m$^{3}$/K produce the best fit. The obtained value of $g_{ei}$ is in good agreement with experimental measurements of $g_{ei}$ in room temperature gold \cite{Probe1,White}. The initial electron temperature corresponds to a corrected energy density in the electron subsystem of $\epsilon\sim 0.17\pm 0.01$~MJ/kg, corresponding to $\eta \sim 0.92$. In the intermediate energy density case, we find that $T_{e}^{0}=8500\pm300$~K and $g_{ei}=5.0\pm 0.5\times10^{16}$~W/m$^{3}$/K produce the best fit, corresponding to a corrected initial energy density for the electron subsystem of $\epsilon \sim 0.21\pm0.01$~MJ/kg, with $\eta \sim 0.57$. Finally, in the high energy density case, we find that the best fit to experimental data comes from values of $T_{e}^{0}=9900\pm300$ K, and $g_{ei}=15.0 \pm 2.0\times 10^{16}$~W/m$^{3}$/K, corresponding to a corrected initial energy density for the electron subsystem of $\epsilon \sim 0.3 \pm 0.01$ MJ/kg, and thus $\eta \sim 0.26$. Our analysis demonstrates a strong dependence of the parameter $\eta$ and the electron-ion equilibration rate on the assumed energy density, i.e., the energy density calculated by taking into consideration purely the reflected and transmitted light (c.f. dashed-black lines in Fig. \ref{fig:finalresult}). 

The bond strength in non-equilibrium gold is predicted to change when the electron temperature is significantly higher than the ion temperature. We check the effect of this on our conclusions by repeating the analysis with two additional interatomic potentials designed to take into account higher temperature electrons \cite{HighTPot}, both of which are validated against finite temperature Kohn-Sham density functional theory \cite{HighTPotValidation}. The first, shown by the dashed-blue lines in Fig. \ref{fig:finalresult}, is calculated for electrons at a temperature of 0.1~eV ($\sim$ 1200~K) and found to be in close agreement with those produced by the interatomic potential developed by Sheng \textit{et al} (A comparison of the properties of these two interatomic potentials can be found in Table \ref{tab:sheng}). Here, we find there to be a ~11\% difference between the best fit results produced by the potential developed by Sheng \textit{et al} and that developed by Norman \textit{et al} \cite{Sheng2011,HighTPot}. For the second potential, calculated for an electron temperature of 1.5~eV ($\sim$17400~K), we only consider results in the high energy density case. Here, we find $\eta=0.15$, which is less than that of the Sheng \textit{et al} case, corresponding to an even more substantial reduction in absorbed energy (c.f. square-red point in Fig. \ref{fig:finalresult}). Fig. S1 and Fig. S2 in the supplemental material depict the full results of this analysis.
 
We find, across multiple potentials, an $\eta$ parameter between 0.2 and 1 that scales strongly with assumed energy density. Our result qualitatively agrees with previous work, which measured a value of 0.5 for similar conditions \cite{Szymon2013}. In this work, they suggest the origin of the lost energy is due to energy dissipation into the supporting grid by ballistic electrons or electron ejection from the target rear \cite{Szymon2013}. Other studies suggest that the temporary formation of a electron sheath around the surface of the target \cite{ProbingWDM} accounts for a decrease in the energy density on a timescale commensurate with the decay of the Laue peaks \cite{ThermionicEmission,WDMCreatedByIso}. Our scaling with assumed energy density, which is proportional to laser fluence for constant target thickness, corroborates this conclusion. We note that this interpretation makes the $\eta$ parameter specific to the target and experimental geometry and thus complicates the interpretation of such experiments. 

\subsection{Spatially-Resolved Diffraction Patterns}

\begin{figure*}[tb!]
    \centering
    \includegraphics[width=0.8\linewidth]{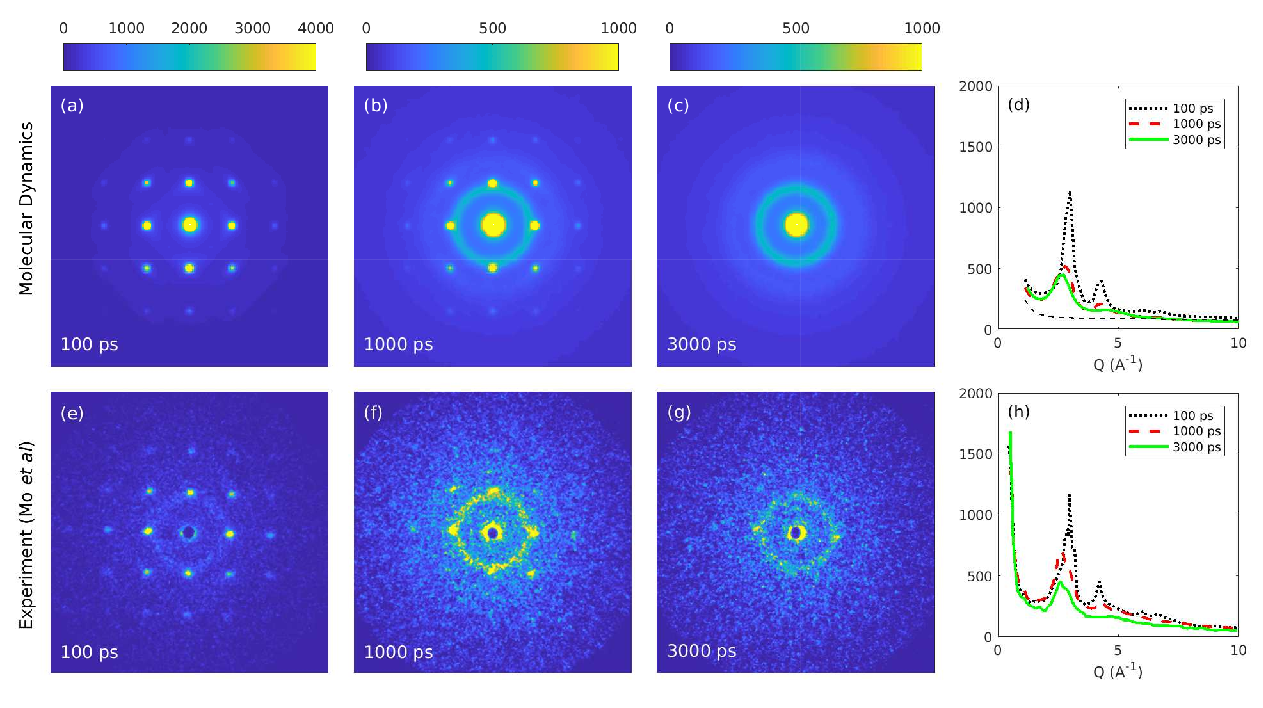}
    \caption{Comparison of simulated and measured spatially-resolved diffraction pattern data for the low energy density case. (a-c) Spatially-resolved data from the simulation for 100~ps, 1000~ps, and 3000~ps. (e-g) Spatially-resolved data obtained by Mo \emph{et al}\cite{Mo2018}. (d) Angularly-resolved diffraction data obtained from azimuthal integrals of the data shown in the panels (a-c). (h) Angularly-resolved diffraction data obtained from azimuthal integrals of the data shown in the panels (e-g).}
    \label{fig:scattering018}
\end{figure*}

\begin{figure*}[tb]
    \centering
    \includegraphics[width=0.8\textwidth]{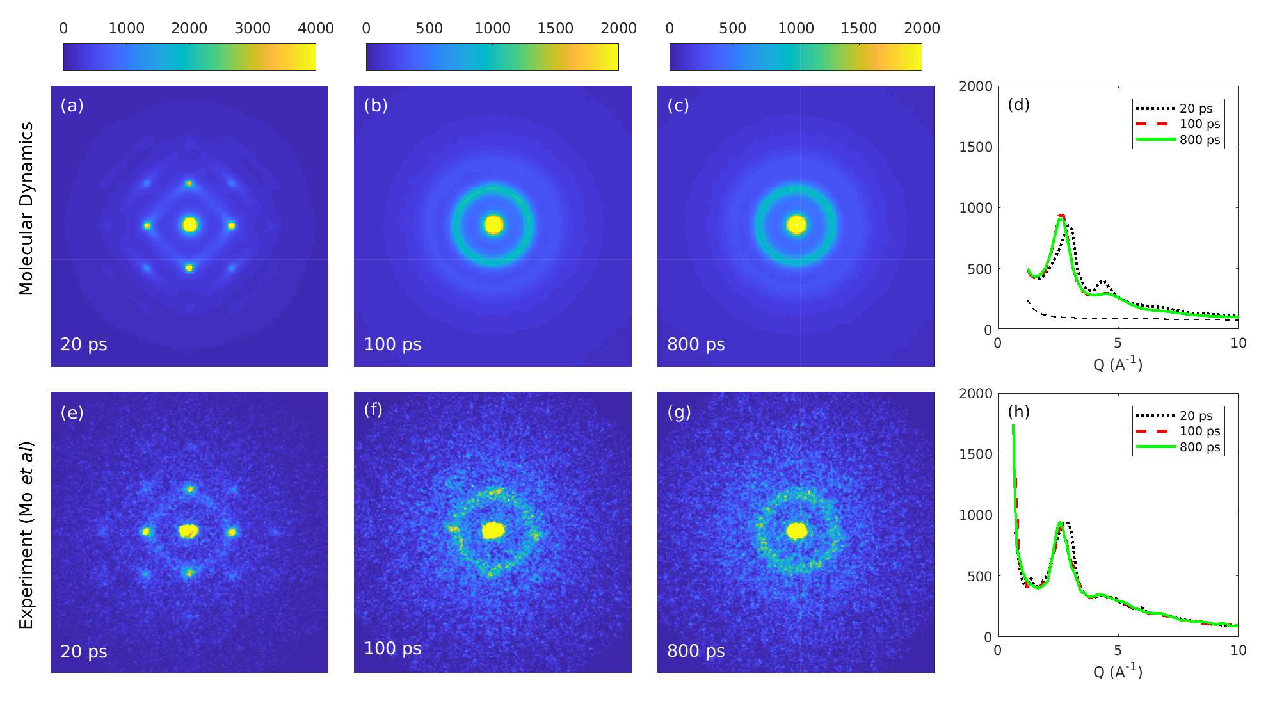}
    \caption{Same as Fig. \ref{fig:scattering018}, but for the intermediate energy density case and times of 20~ps, 100~ps, and 800~ps.}
    \label{fig:scattering036}
\end{figure*}

\begin{figure*}[tb]
    \centering
    \includegraphics[width=0.8\textwidth]{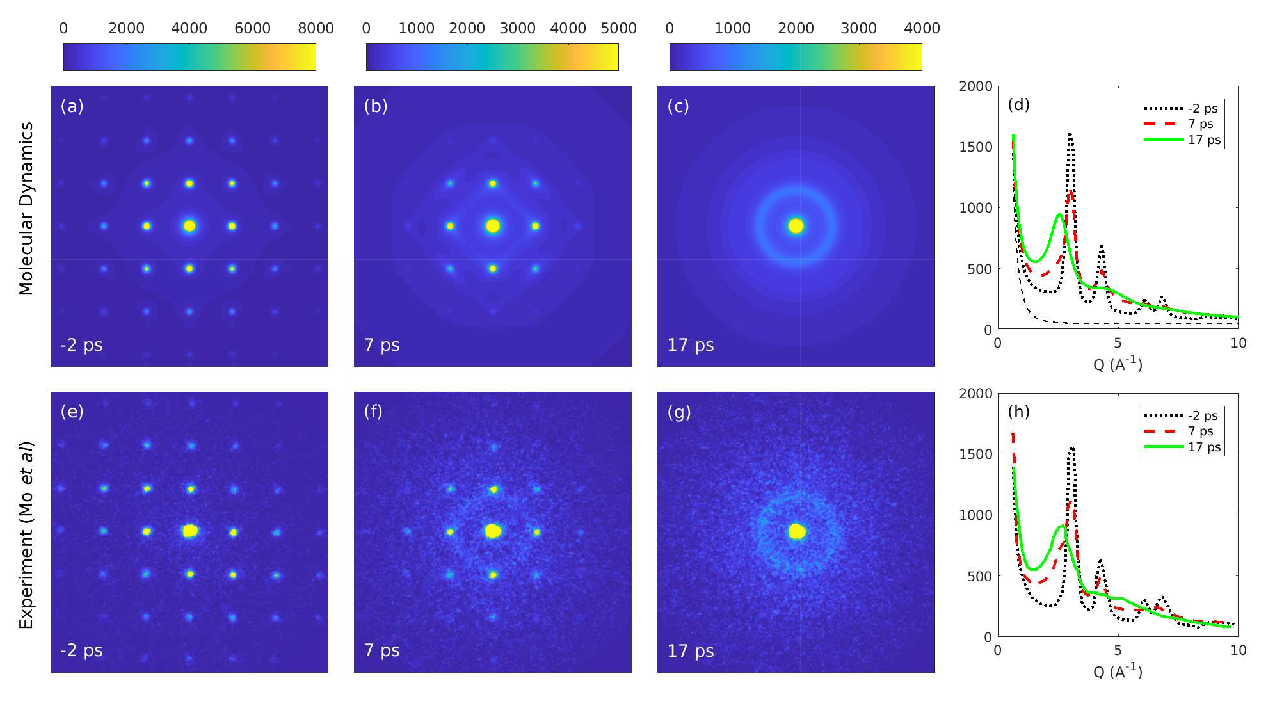}
    \caption{Same as Fig. \ref{fig:scattering018}, but for the high energy density case and the times of -2~ps, 7~ps, and 17~ps.}
    \label{fig:scattering117}
\end{figure*}
For each of the three energy density cases, we took the best simulation achieved with a constant $g_{ei}$, and predicted spatially-resolved synthetic diffraction patterns. For the first $50$~ps, these were created by Fourier Transforming the atomic positions taken from simulations using the larger geometry and taking a $Q_x$-$Q_y$ slice of reciprocal space. In each case, we confirmed that these larger simulations reproduced the Laue peak decay seen in the smaller simulations. The diffraction patterns, shown in Figs. \ref{fig:scattering018}-\ref{fig:scattering117}, correspond to the three energy density cases.The timescales of the experimental diffraction pattern measurements necessitated the use of the smaller simulation geometry in producing Figs. \ref{fig:scattering018}-\ref{fig:scattering117} for times above $50$~ps. Also shown, for comparison, are the experimentally obtained diffraction patterns. 

In order for a direct comparison with the measured spatially-resolved electron scattering data of Mo \emph{et al}, the simulated structure factor was multiplied by the electron scattering form factor, which was calculated using the Mott–Bethe formula \cite{ElecForm} and utilized tabulated X-ray elastic scattering cross-sections \cite{xrayform}. We found negligible difference with using either a cold or singularly ionized form factor, which exhibit little difference above $\sim2~$\AA$^{-1}$. In addition, a background term of the form \cite{inelastic1,inelastic2},
\begin{equation}
\frac{\mathrm{d} \sigma_{\mathrm{inel}}}{\mathrm{d} \Omega}=\frac{A}{Q^{4}}\left[1-\frac{1}{\left(1+B~Q^{2}\right)^{2}}\right] + mQ+c,
\end{equation} 
was introduced to account for inelastic scattering. The coefficients, $A=350$ and $B=1.3$, were kept constant for all comparisons, whereas small variations were required in the linear component between shots at different laser intensities. In each case the background term is plotted as a dashed line in the figures and represents a small component of the overall signal. Finally, the entire image was convoluted with a pseudo-Voigt profile that represented the spatial profile of the electron beam \cite{voigt} (FWHM$_\mathrm{Gauss}$=17~meV, FWHM$_\mathrm{Lorentz}$=6.5~meV).

Initially, we found that the diffraction peaks from the solid gold were an order of magnitude larger than the liquid scattering signal, an effect not seen in the experimental data where the difference is closer to a factor of two. We attribute this discrepancy to misalignment and beam divergence in the experiment that is not present in the simulations. Thus, we average over a small $\Delta Q_z$, reducing the intensity of the solid diffraction peaks while leaving the liquid diffraction peaks, which are broad in reciprocal space, unchanged. We choose $\Delta Q_z$, which remains constant in our modelling, by matching the spatially-resolved scattering signal at both early (t=0~ps) and late (t=17~ps) times in the high-energy-density case. With these changes we are able to match, in each of the three energy density cases, the angularly-resolved line-outs obtained in the experiment. It is important to note that the applied modifications to the raw data, used to obtain the angularly-resolved data, affect only the qualitative comparisons provided in Fig. \ref{fig:scattering018} - Fig. \ref{fig:scattering117}, and not the intensity decay curves used in Fig. \ref{fig:constanteq}.

\subsection{Heterogeneous/Homogeneous Melting}

\begin{figure}[tb!]
    \centering
    \includegraphics[width=0.79\linewidth]{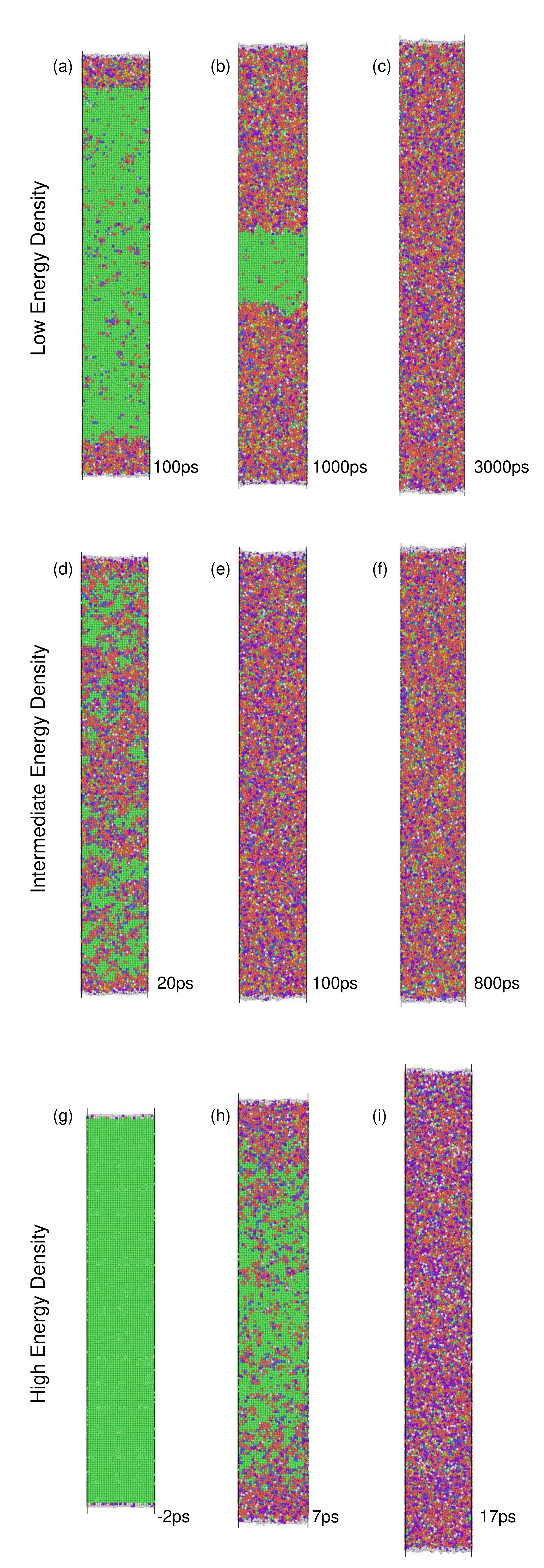}
    \caption{Visualization of gold melting in the low (a-c), intermediate (d-f), and high (g-i) energy density cases via the technique of polyhedral template matching \cite{Poly} in the OVITO visualization software \cite{OvitoRef}. For each energy density case, the time steps corresponding to the diffraction patterns shown in Figs. \ref{fig:scattering018}, \ref{fig:scattering036}, \ref{fig:scattering117} are presented. The color green identifies the FCC lattice type, whereas the non-green colors identify liquid present in the sample. These results were produced utilizing the smaller simulation geometry.}
    \label{fig:SimMelt}
\end{figure}

In our analysis of the melt behavior of the sample, we utilize the smaller simulation geometry. This was due to the lack of available computational resources that would allow us to perform simulations, in the large simulation geometry, that capture the full duration of the low and intermediate energy density cases ($3000$~ps and $800$~ps respectively). We were, however, able to compare the small and large geometries, in all three energy density cases, out to $50$~ps. In all energy density cases, we find the melt behavior and the expansion rates of the large and small geometry samples to be equivalent out to $50$~ps.

From the time for the total disappearance of the (200) peak, the three energy density cases were thought to represent the incomplete, heterogeneous, and homogeneous melting processes respectively \cite{Mo2018}. Based on this analysis, they concluded a threshold for complete melting (assuming total absorption of the laser energy, $\eta=1$) of $\sim$0.25~MJ/kg, and a transition between heterogeneous and homogeneous melting of $\sim$0.38~MJ/kg. Of note, both of these values were higher than those predicted by molecular dynamics simulations employing the interatomic potential of Zhou \emph{et al} \cite{Lin2006, Zhou}. 

In our analysis, utilizing the interatomic potential developed by Sheng \textit{et al} \cite{Sheng2011}, we are able to match the entire decay of the experimentally measured Laue peak. Using the OVITO visualization tool \cite{OvitoRef} together with polyhedral template matching \cite{Poly} we can directly assess the structure and melting process in each of our cases (c.f. Fig. \ref{fig:SimMelt}). In contrast to the original work, we find that the low energy case exhibits heterogeneous melting, with a clear propagation of the melt front (see Fig \ref{fig:SimMelt}(a-c)) and no nucleation of melt inside the target. The two higher energy densities both exhibit homogeneous melting, with the intermediate case appearing to be on the cusp of the two regimes, as determined from non-uniform melt regions inside the target. Taking this into consideration, and the lower predicted energy density from the $\eta$ parameter, we find a total energy density threshold for complete melting below 0.22~MJ/kg (which includes the 0.17~MJ/kg present in the electron subsystem and the 0.05~MJ/kg already present in our 300~K gold sample). This value is in agreement with the expected value of 0.22~MJ/kg \cite{Lin2010}. For the transition between homogeneous and heterogeneous melting, we find a value that lies between 0.22~MJ/kg and 0.26~MJ/kg. With the introduction of the $\eta$ parameter, we find that the complete melting threshold, and heterogeneous-to-homogeneous melting boundaries, are both lower than previously thought\cite{Mo2018}. 

Finally, we note that despite achieving excellent agreement with the experimental decay curves, we are not able to reproduce the low-intensity, long-lived, diffraction features present in the experiment (c.f. Fig. \ref{fig:scattering036}(b) and (f)). We find that the intermediate energy cases melt within 100~ps, whereas the original work concludes that it does not melt until 800~ps. To investigate if this was a fault of the equilibration rate, we manually adjusted the value of $g_{ei}$ throughout the simulation to best match the experimental results; we were unable to match both the early time Laue peak decay and late-time long-lived diffraction patterns. In the simulation, this may highlight a deficiency in either the TTM model or the interatomic potential. In the experiment, this could be attributed to inhomogeneous heating.

\section{Temperature-Dependent Electron-Ion Equilibration Rates}

\begin{figure}[tb!]
    \centering
   \includegraphics[width=\linewidth]{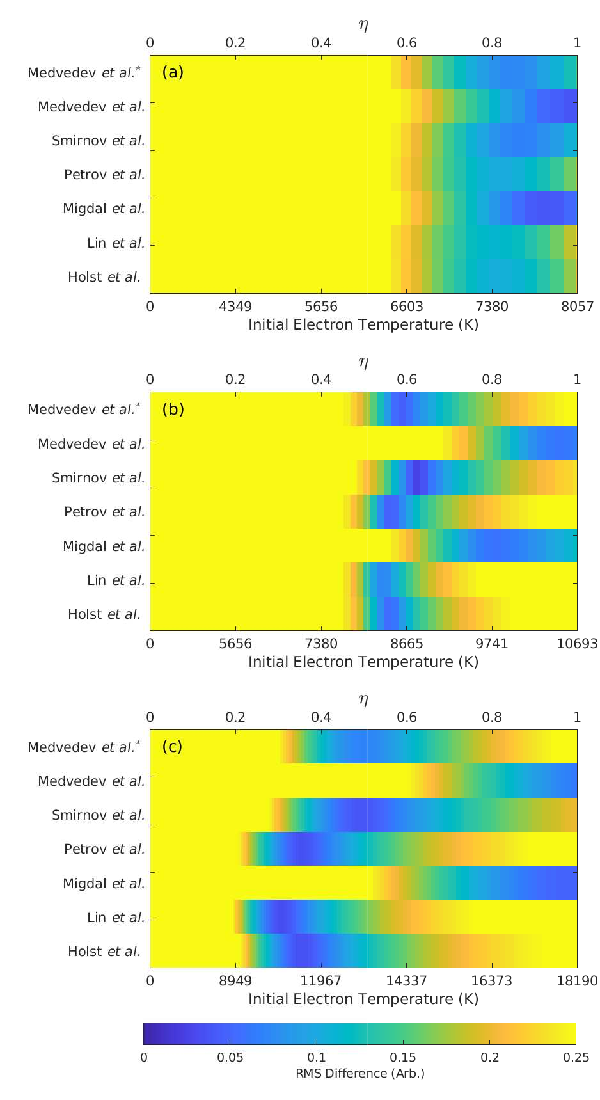}
    \caption{Comparison between experimentally obtained Laue decay curves \cite{Mo2018} and those produced through molecular dynamics simulations utilizing temperature-dependent electron-ion equilibration rates \cite{Holst2018,Lin2008,Med2020,Smirnov2020,Migdal2013,Petrov2013} for the low (a), intermediate (b), and high (c) energy density cases. Here, Medvedev \textit{et al}* is used to denote a $g_{ei}(T)$ prediction that is both electron and ion temperature-dependent \cite{Med2020}. For all three energy density cases, the RMS difference between simulated and experimental decay curves for a range of initial electron temperatures is shown for each of the employed $g_{ei}(T)$ models. In our analysis, the RMS difference of the (220) and (420) have been added in quadrature.}
    \label{fig:theories}
\end{figure}

The electron-ion equilibration rate, $g_{ei}$, is predicted to scale strongly with both electron and ion temperature. Thus, to supplement our study, we employ seven models of temperature-dependent equilibration rates\cite{Holst2018,Lin2008,Med2020,Smirnov2020,Migdal2013,Petrov2013}. This set of simulations was performed under the same specifications outlined in section 2 and used the smaller simulation geometry. As in section 3, we use the decay of the (220) and (420) Laue diffraction peaks in order to draw comparisons between simulation and the experimental results measured by Mo \textit{et al}. In each energy density case, we consider only a temperature range corresponding to $\eta$ of $\sim 0 \leq \eta \leq 1$, so as to only consider physically possible values. The performance of each model can be seen in Fig. \ref{fig:theories}. In the low energy density case, shown in Fig. \ref{fig:theories}(a), we find that the Medvedev \textit{et al} model, calculated for a constant ion temperature of 300~K, and the Migdal \textit{et al} models performed the best. For the intermediate energy density case, shown in Fig. \ref{fig:theories}(b), we find that the model developed by Smirnov \textit{et al} had the lowest RMS difference. Finally, in the high energy density case, shown in Fig. \ref{fig:theories}(c), we find that the models developed by Lin \textit{et al}, Holst \textit{et al}, and Petrov \textit{et al}, perform the best.  Considering all three energy density cases simultaneously, the model by Smirnov \textit{et al} performs the best. 

The success of these theoretical models is contingent on a similar inclusion of energy loss as was found in the constant $g_{ei}$ description. Using the best fit value of $T_{e}^{0}$, we can calculate the $\eta$ parameter for each theoretical model and each energy density case. The evolution of the $\eta$ factor for both the constant $g_{ei}$ case and the temperature-dependent $g_{ei}$ models is shown in Fig. \ref{fig:finalresult}(a). The electron temperature-dependent models display a similar trend in $\eta$ as the constant case, except for the solely electron temperature-dependent model developed by Medvedev \textit{et al}, calculated for a constant ion temperature of 300~K, and the model developed by Migdal \textit{et al} \cite{Med2020,Migdal2013}. Excluding these outliers, based on the fact they predict either energy gain or no energy loss in the system, we calculate the average $\eta$ across all of our tested models. We find values of $\bar \eta \sim 0.83$, $\bar \eta \sim 0.57$, and $\bar \eta \sim 0.40$ for the low, intermediate, and high energy density cases respectively; very close to the constant $g_{ei}$ case.

In Fig. \ref{fig:finalresult}(b), as a method of comparing models of $g_{ei}$ with different functional dependencies, we plot the average value of $g_{ei}$ in the best fit simulation for each theoretical model, as well as the constant $g_{ei}$ cases discussed in section 3. In each simulation, we average the value of $g_{ei}$ out until the time at which the decrease in the Laue decay curves stops; in the low, intermediate, and high energy density cases, these times corresponded to 50~ps, 25~ps, and 10~ps respectively. In Fig. \ref{fig:finalresult}(a) and Fig. \ref{fig:finalresult}(b) the best fit results for the constant and temperature-dependent $g_{ei}$ models are in close agreement with one another for the low and intermediate energy density cases. However, they diverge in the high energy density case. This discrepancy could be due to bond softening, which is predicted to occur in gold, at its equilibrium volume, for electron temperatures $T_{e}>9000$~K (or, an energy density of $\sim$ 0.24~MJ/kg) \cite{Szymon2013}. Our results suggest that in the high energy density case the initial electron temperature of the sample is expected to be above $T_{e}^{0}\sim$9000~K.

\section{Conclusion}

We benchmark the molecular dynamics simulations of the ultrafast excitation of thin gold films against experimental MeV electron diffraction results published by Mo \emph{et al}. We carry out this analysis using a constant electron-ion coupling rate, seven prominent temperature-dependent electron-ion coupling rate models, as well as for three different interatomic potentials. Using the Laue peak decay as a metric, and treating the initial energy density and the electron-ion equilibration rate as free parameters, we find a strong dependence of the $\eta$ parameter on initial energy density, with increased energy loss at higher energy densities (and higher laser fluence), which is in good agreement with the work of Daraszewicz \textit{et al} \cite{Szymon2013}. We believe the need for an additional energy loss pathway, and the presented scaling with increased fluence, to be reliably established. Ultimately this ambiguity needs to be resolved via direct measurement of the sub-system temperatures. 

\begin{acknowledgments}
The authors acknowledge M. Z. Mo for providing the experimental data that was used throughout this work. This work was funded in part by the US Department of Energy, National Nuclear Security Administration (NNSA) (Award No. DE- NA0004039). This material is based upon work supported by the National Science Foundation under Grant Number 2045718.  The authors acknowledge the support of Research \& Innovation and the Office of Information Technology at the University of Nevada, Reno for computing time on the Pronghorn High Performance Computing Cluster. J. M. Molina acknowledges the funding and support from the Nevada NSF-EPSCoR NEXUS Award and the McNair Scholars Program at the University of Nevada, Reno. 
\end{acknowledgments}

\end{document}